\documentstyle[12pt,epsf]{article}
\newlength{\pubnumber} 

\catcode`\@=11
\@addtoreset{equation}{section}

\def\section{\@startsection{section}{1}{\z@}{3.5ex plus 1ex minus .2ex}
 {2.3ex plus .2ex}{\large\bf}}
\def\subsection{\@startsection{subsection}{2}{\z@}{2.3ex plus .2ex}
 {2.3ex plus .2ex}{\bf}}

\begin{document}

\begin{titlepage}
\samepage{
\setcounter{page}{0}
\rightline{TPI--MINN--99/32}
\rightline{UMN--TH--1806--99}
\rightline{\tt hep-ph/9906345}
\rightline{June 1999}
\vfill
\begin{center}
 {\Large \bf Probing the Desert with Ultra--Energetic Neutrinos
from the Sun and the Earth}
\vfill
\vspace{.25in}
 {\large Alon E. Faraggi, Keith A. Olive $\,$and$\,$ Maxim Pospelov\\}
\vspace{.25in}
 {\it  Department of Physics,
              University of Minnesota, Minneapolis, MN  55455, USA\\}
\end{center}
\vfill
\begin{abstract}
  {\rm
Realistic superstring models generically give rise to exotic
matter states, which arise due to the ``Wilson--line''
breaking of the non--Abelian unifying gauge symmetry.
Often such states are protected by a gauge or local discrete
symmetry and therefore may be stable or meta--stable.
We study the possibility of a flux of high energy neutrinos
coming from the sun and the earth due to the annihilation 
of such exotic string states.
We also discuss the expected flux for other heavy stable
particles -- like the gluino LSP. 
We comment that the detection of ultra--energetic neutrinos from the
sun and the earth imposes model independent constraints on the 
high energy cutoff, as for example in the recently
entertained TeV scale Kaluza--Klein theories.
We therefore propose that improved experimental resolution
of the energy of the muons in neutrino detectors together
with their correlation with neutrinos from the sun and the center of the earth
will serve as a probe of the desert in Gravity Unified Theories.
}
\end{abstract}
\vfill
\smallskip}
\end{titlepage}

\setcounter{footnote}{0}

\def\beq{\begin{equation}}
\def\eeq{\end{equation}}
\def\beqn{\begin{eqnarray}}
\def\la{\label}
\def\eeqn{\end{eqnarray}}
\def\Tr{{\rm Tr}\,}
\def\KM{{Ka\v{c}-Moody}}

\def\ie{{\it i.e.}}
\def\etc{{\it etc}}
\def\eg{{\it e.g.}}
\def\half{{\textstyle{1\over 2}}}
\def\third{{\textstyle {1\over3}}}
\def\quarter{{\textstyle {1\over4}}}
\def\m{{\tt -}}
\def\p{{\tt +}}

\def\rep#1{{\bf {#1}}}
\def\slash#1{#1\hskip-6pt/\hskip6pt}
\def\slk{\slash{k}}
\def\GeV{\,{\rm GeV}}
\def\TeV{\,{\rm TeV}}
\def\y{\,{\rm y}}
\def\SM{Standard-Model }
\def\SUSY{supersymmetry }
\def\SSM{supersymmetric standard model}
\def\vev#1{\left\langle #1\right\rangle}
\def\l{\langle}
\def\r{\rangle}

\def\Htw{{\tilde H}}
\def\chibar{{\overline{\chi}}}
\def\qbar{{\overline{q}}}
\def\ibar{{\overline{\imath}}}
\def\jbar{{\overline{\jmath}}}
\def\Hbar{{\overline{H}}}
\def\Qbar{{\overline{Q}}}
\def\abar{{\overline{a}}}
\def\alphabar{{\overline{\alpha}}}
\def\betabar{{\overline{\beta}}}
\def\tautwo{{ \tau_2 }}
\def\calF{{\cal F}}
\def\calP{{\cal P}}
\def\calN{{\cal N}}
\def\smallmatrix#1#2#3#4{{ {{#1}~{#2}\choose{#3}~{#4}} }}
\def\bone{{\bf 1}}
\def\V{{\bf V}}
\def\N{{\bf N}}
\def\bQ{{\bf Q}}
\def\t#1#2{{ \Theta\left\lbrack \matrix{ {#1}\cr {#2}\cr }\right\rbrack }}
\def\C#1#2{{ C\left\lbrack \matrix{ {#1}\cr {#2}\cr }\right\rbrack }}
\def\tp#1#2{{ \Theta'\left\lbrack \matrix{ {#1}\cr {#2}\cr }\right\rbrack }}
\def\tpp#1#2{{ \Theta''\left\lbrack \matrix{ {#1}\cr {#2}\cr }\right\rbrack }}
\def\ul#1{$\underline{#1}$}
\def\bE#1{{E^{(#1)}}}
\def\IZ{\relax{\bf Z}}\def\IC{\relax{\bf C}}
\def\IR{\relax{\rm I\kern-.18em R}}
\def\lamb{\lambda}
\def\fc#1#2{{#1\over#2}}
\def\hx#1{{\hat{#1}}}
\def\Gh{\hat{\Gamma}}
\def\subsubsec#1{\noindent {\it #1} \br}
\def\WP{{\bf WP}}
\def\gn{\Gamma_0}
\def\bgn{{\bar \Gamma}_0}
\def\Ds{\Delta^\star}
\def\abstract#1{
\vskip .5in\vfil\centerline
{\bf Abstract}\penalty1000
{{\smallskip\ifx\answ\bigans\leftskip 2pc \rightskip 2pc
\else\leftskip 5pc \rightskip 5pc\fi
\noindent\abstractfont \baselineskip=12pt
{#1} \smallskip}}
\penalty-1000}
\def\us#1{\underline{#1}}
\def\hth/#1#2#3#4#5#6#7{{\tt hep-th/#1#2#3#4#5#6#7}}
\def\nup#1({Nucl.\ Phys.\ $\us {B#1}$\ (}
\def\plt#1({Phys.\ Lett.\ $\us  {B#1}$\ (}
\def\cmp#1({Commun.\ Math.\ Phys.\ $\us  {#1}$\ (}
\def\prp#1({Phys.\ Rep.\ $\us  {#1}$\ (}
\def\prl#1({Phys.\ Rev.\ Lett.\ $\us  {#1}$\ (}
\def\prv#1({Phys.\ Rev.\ $\us  {#1}$\ (}
\def\mpl#1({Mod.\ Phys.\ Let.\ $\us  {A#1}$\ (}
\def\ijmp#1({Int.\ J.\ Mod.\ Phys.\ $\us{A#1}$\ (}
\def\br{\hfill\break}\def\ni{\noindent}
\def\mbr{\hfill\break\vskip 0.2cm}
\def\cx#1{{\cal #1}}\def\al{\alpha}\def\IP{{\bf P}}
\def\ov#1#2{{#1 \over #2}}
\def\b{{\bf b}}
\def\S{{\bf S}}
\def\X{{\bf X}}
\def\I{{\bf I}}
\def\mb{{\mathbf b}}
\def\mS{{\mathbf S}}
\def\mX{{\mathbf X}}
\def\mI{{\mathbf I}}
\def\balpha{{\mathbf \alpha}}
\def\bbeta{{\mathbf \beta}}
\def\bgamma{{\mathbf \gamma}}
\def\bxi{{\mathbf \xi}}
 
\def\ul#1{$\underline{#1}$}
\def\bE#1{{E^{(#1)}}}
\def\IZ{\relax{\bf Z}}\def\IC{\relax{\bf C}}
\def\IR{\relax{\rm I\kern-.18em R}}
\def\lam{\lambda}
\def\fc#1#2{{#1\over#2}}
\def\hx#1{{\hat{#1}}}
\def\Gh{\hat{\Gamma}}
\def\subsubsec#1{\noindent {\it #1} \br}
\def\WP{{\bf WP}}
\def\gn{\Gamma_0}
\def\bgn{{\bar \Gamma}_0}
\def\Ds{\Delta^\star}
\def\abstract#1{
\vskip .5in\vfil\centerline
{\bf Abstract}\penalty1000
{{\smallskip\ifx\answ\bigans\leftskip 2pc \rightskip 2pc
\else\leftskip 5pc \rightskip 5pc\fi
\noindent\abstractfont \baselineskip=12pt
{#1} \smallskip}}
\penalty-1000}
\def\us#1{\underline{#1}}
\def\hth/#1#2#3#4#5#6#7{{\tt hep-th/#1#2#3#4#5#6#7}}
\def\nup#1({Nucl.\ Phys.\ $\us {B#1}$\ (}
\def\plt#1({Phys.\ Lett.\ $\us  {B#1}$\ (}
\def\cmp#1({Commun.\ Math.\ Phys.\ $\us  {#1}$\ (}
\def\prp#1({Phys.\ Rep.\ $\us  {#1}$\ (}
\def\prl#1({Phys.\ Rev.\ Lett.\ $\us  {#1}$\ (}
\def\prv#1({Phys.\ Rev.\ $\us  {#1}$\ (}
\def\mpl#1({Mod.\ Phys.\ Let.\ $\us  {A#1}$\ (}
\def\ijmp#1({Int.\ J.\ Mod.\ Phys.\ $\us{A#1}$\ (}
\def\br{\hfill\break}\def\ni{\noindent}
\def\mbr{\hfill\break\vskip 0.2cm}
\def\cx#1{{\cal #1}}\def\al{\alpha}\def\IP{{\bf P}}
\def\ov#1#2{{#1 \over #2}}
\def\ga{\mathrel{\raise.3ex\hbox{$>$\kern-.75em\lower1ex\hbox{$\sim$}}}}
\def\la{\mathrel{\raise.3ex\hbox{$<$\kern-.75em\lower1ex\hbox{$\sim$}}}}
\def\deg{\hbox{${}^\circ$}}     


\def\inbar{\,\vrule height1.5ex width.4pt depth0pt}

\def\IC{\relax\hbox{$\inbar\kern-.3em{\rm C}$}}
\def\IQ{\relax\hbox{$\inbar\kern-.3em{\rm Q}$}}
\def\IR{\relax{\rm I\kern-.18em R}}
 \font\cmss=cmss10 \font\cmsss=cmss10 at 7pt
\def\IZ{\relax\ifmmode\mathchoice
 {\hbox{\cmss Z\kern-.4em Z}}{\hbox{\cmss Z\kern-.4em Z}}
 {\lower.9pt\hbox{\cmsss Z\kern-.4em Z}}
 {\lower1.2pt\hbox{\cmsss Z\kern-.4em Z}}\else{\cmss Z\kern-.4em Z}\fi}

\def\AEF{A.E. Faraggi}
\def\KRD{K.R. Dienes}
\def\JMR{J. March-Russell}
\def\MEP{M.E. Pospelov}
\def\NPB#1#2#3{{\it Nucl.\ Phys.}\/ {\bf B#1} (19#2) #3}
\def\PLB#1#2#3{{\it Phys.\ Lett.}\/ {\bf B#1} (19#2) #3}
\def\PRD#1#2#3{{\it Phys.\ Rev.}\/ {\bf D#1} (19#2) #3}
\def\PRL#1#2#3{{\it Phys.\ Rev.\ Lett.}\/ {\bf #1} (19#2) #3}
\def\PRT#1#2#3{{\it Phys.\ Rep.}\/ {\bf#1} (19#2) #3}
\def\MODA#1#2#3{{\it Mod.\ Phys.\ Lett.}\/ {\bf A#1} (19#2) #3}
\def\IJMP#1#2#3{{\it Int.\ J.\ Mod.\ Phys.}\/ {\bf A#1} (19#2) #3}
\def\nuvc#1#2#3{{\it Nuovo Cimento}\/ {\bf #1A} (#2) #3}
\def\etal{{\it et al,\/}\ }

\hyphenation{su-per-sym-met-ric non-su-per-sym-met-ric}
\hyphenation{space-time-super-sym-met-ric}
\hyphenation{mod-u-lar mod-u-lar--in-var-i-ant}
\newcommand{\be}{\begin{equation}}
\newcommand{\ee}{\end{equation}}
\newcommand{\als}{\mbox{$\alpha_{s}$}}
\newcommand{\s}{\mbox{$\sigma$}}
\newcommand{\lm}{\mbox{$\mbox{ln}(1/\alpha)$}}
\newcommand{\bi}[1]{\bibitem{#1}}
\newcommand{\fr}[2]{\frac{#1}{#2}}
\newcommand{\sv}{\mbox{$\vec{\sigma}$}}
\newcommand{\gm}{\mbox{$\gamma_{\mu}$}}
\newcommand{\Pm}{\mbox{$P_{\mu}$}}
\newcommand{\Pn}{\mbox{$P_{\nu}$}}
\newcommand{\Pa}{\mbox{$P_{\alpha}$}}
\newcommand{\ph}{\mbox{$\hat{p}$}}
\newcommand{\Ph}{\mbox{$\hat{P}$}}
\newcommand{\qh}{\mbox{$\hat{q}$}}
\newcommand{\kh}{\mbox{$\hat{k}$}}
\newcommand{\Le}{\mbox{$\fr{1+\gamma_5}{2}$}}
\newcommand{\R}{\mbox{$\fr{1-\gamma_5}{2}$}}
\newcommand{\GD}{\mbox{$\tilde{G}$}}
\newcommand{\gf}{\mbox{$\gamma_{5}$}}
\newcommand{\om}{\mbox{$\omega$}}
\newcommand{\Ima}{\mbox{Im}}
\newcommand{\Rea}{\mbox{Re}}


\setcounter{footnote}{0}
\section{Introduction}
The spectacular confirmation of the Standard Model at LEP/SLC
and other high energy colliders \cite{review},
as well as the proton longevity
and suppression of left--handed neutrino masses, provide strong
support for the grand desert scenario and unification.
Indeed, the question of how to test this hypothesis
in present and future experiments constitutes much
of today's activity in high energy physics.
Direct signatures at LHC/NLC will be able to probe
the desert only up to a few TeV. On the other hand,
it has been suggested that all present day
experimental data including ultra--high energy
cosmic rays above the Greisen--Zatsepin--Kuzmin
cutoff, cannot rule out the possibility that the
fundamental scale is near the TeV scale \cite{add}. 

In this paper we examine the potential of probing the desert
using ultra energetic neutrinos from the sun and the earth. 
Such ultra--energetic neutrinos arise if the dark matter
in the galactic halo is composed of heavy long--lived matter,
that is trapped in the sun and the earth and subsequently
annihilates into neutrinos \cite{sos}. The magnitude
of the expected signal depends on the competition between the trapping
and annihilation rates of specific dark matter candidates,
which in turn depends on the specific properties of
such states. The well known case which has 
been studied extensively in the past is 
the trapping and annihilation of stable neutralinos
which arise in supersymmetric theories with
conserved R--parity \cite{sos,kamion,berg}. 

High energy ($E >$ 1 GeV) 
neutrinos from the sun and the earth can only be produced by the decay or
annihilation of a heavy particle.
For the simple reason that, unlike charged matter, neutrinos
cannot be accelerated, the observation of
ultra--energetic neutrinos whose production can be
definitely correlated with the sun or the center of the earth, will 
signal new physics at the high energy scale.  
In addition, such an observation would
also impose a model independent constraint on the fundamental
cutoff scale. 

Superstring models generically give rise to exotic matter
states which arise because of the breaking of the non--Abelian unifying
gauge symmetry, $G$, by Wilson--lines \cite{yut,ww,ccf}.
In many respects the unifying gauge symmetry
is similar to the gauge group of four dimensional grand unification
and the Wilson lines are similar to the Higgs bosons in the
adjoint representation. However, there are some notable differences.
The eigenvalues of the Wilson lines are quantized while the
eigenvalues of the Higgs in the adjoint representation are continuous.
Another important difference is that the breaking of the gauge
symmetries by Wilson lines results in massless states that
do not fit into multiplets of the original unbroken
gauge symmetry. We refer to such states generically as
exotic ``Wilsonian'' matter states. This is an important
property as it may result in conserved quantum numbers that will
indicate the stability of these massless ``Wilsonian'' states.
The simplest example of this phenomenon is the existence
of states with fractional electric charge in the massless
spectrum of superstring models \cite{ww,eln,huet,fcp}.
Such states are stable due to electric charge conservation.
As there exist strong constraints on their masses and abundance,
states with fractional electric charge must be diluted away or
extremely massive.
The same ``Wilson line''
breaking mechanism, which produces matter with fractional electric
charge, is also responsible for the existence of states which
carry the ``standard'' charges under the Standard Model gauge
group but which carry fractional charges under a different subgroup
of the unifying gauge group. For example, if the group $G$ is
$SO(10)$ then the ``Wilsonian'' states may carry non--standard
charges under the $U(1)_{Z^\prime}$ symmetry, which is embedded
in $SO(10)$ and is orthogonal to $U(1)_Y$.
Such states can therefore be long--lived if the $U(1)_{Z^\prime}$ gauge
symmetry remains unbroken down to low energies, or if some
residual local discrete symmetry is left unbroken after 
$U(1)_{Z^\prime}$ symmetry breaking.
The existence of heavy stable ``Wilsonian''
matter can be argued to be a ``smoking gun'' of string unification.

There are several examples of ``Wilsonian matter'' in the literature.
The uniton is a dark matter candidate which arises in
realistic  heterotic string models \cite{ccf}. We show that the limits due
to energetic neutrinos from the sun and the earth are far more restrictive
than the constraints arising from over-closure
of the universe in the case of the uniton due to its strong interaction
with matter. As another example of a strongly interacting stable particle,
we also consider the constraints on recent proposals of gluino--LSP
\cite{raby} (which has not been suggested as a dark matter
candidate), and show that the constraints imposed from annihilation into
energetic neutrinos in the sun are sufficiently
restrictive to rule out this possibility. Other astrophysical constraints
applicable to the gluino LSP were considered in \cite{mt}. Other
candidates we consider are the {\it cryptons} of ref.
\cite{cryptons} which are fractionally charged states confined by a
hidden non-Abelian gauge symmetry, and other ``Wilsonian matter'' states
which may consist  of free fractionally charged states or Standard Model
neutral states with fractional charges under the
$U(1)_{Z^\prime}$ symmetry. We discuss the potential for
discovery of each of these states by energetic neutrinos
from the sun. 

Recently there has been considerable interest in the possibility 
that a very massive, long-lived particle can account for both dark matter
and the observed ultra-high energy cosmic rays (above the GZK cutoff
\cite{GZK}) through its decays \cite{ber}. For example, a particle with a
mass of $10^{13-14}$ GeV with a lifetime (depending on the relic density)
greater than the age of the Universe could possibly explain these
phenomena, although the more detailed computation in \cite{sb} indicates
that the particle mass should be about $10^{11-12}$ GeV. The
specific case of cryptons in this context was considered in
\cite{ben,sb}. Though one would not expect the lifetime for such a massive
particle to be sufficiently long, discrete symmetries have been proposed
to achieve the required stability
\cite{yan}. The light gluino candidates of \cite{raby} are too few in
number to account for the dark matter in the galaxy, nevertheless, they
have been proposed as a possible source for the ultra-high energy cosmic
rays as do the very light gluinos discussed in
\cite{cfk}. The latter can not be tested by annihilation in the sun or 
the earth as they
are too light and would produce neutrinos below threshold.

In what follows, we first describe the types of Wilsonian dark matter
states that can be expected to arise in realistic string theory
formulations and describe the specific candidates we consider. 
We then go on to compute the flux of neutrinos generated in the core of
the sun and the earth and estimate the flux of
upward-going muons which signal the annihilation into neutrinos.
We find that that certain candidates such as the uniton which have
strong interactions can be excluded as dark matter.  As a source of the
ultra-high energy cosmic-rays, the gluino--LSP candidate can also be
excluded.  For other candidates, such the crypton, the best one can do is
place limits on its mass and interaction strength.  

\setcounter{footnote}{0}
\section{Realistic free fermionic models}

The realistic models in the free fermionic formulation are generated by
a basis of boundary condition vectors for all world--sheet fermions
\cite{revamp}-\cite{SLM}.
The basis is constructed in two stages. The first stage consists
of the NAHE set \cite{revamp,SLM},
which is a set of five boundary condition basis
vectors, $\{{{\bf 1},S,b_1,b_2,b_3}\}$. The gauge group after the NAHE set
is $SO(10)\times SO(6)^3\times E_8$ with $N=1$ space--time supersymmetry.
The NAHE set correspond to $Z_2\times Z_2$ orbifold compactification
\cite{foc}. The Neveu--Schwarz sector
corresponds to the untwisted sector, and the sectors
$b_1$, $b_2$ and $b_3$ to the three twisted sectors
of the $Z_2\times Z_2$ orbifold model.
In addition to the gravity and gauge multiplets, the Neveu--Schwarz
sector produces six multiplets in the 10 representation
of $SO(10)$, and several $SO(10)$ singlets transforming under the
flavor $SO(6)^3$ symmetries.
The sectors $b_1$, $b_2$ and $b_3$
produce 48 spinorial 16's of $SO(10)$,
sixteen each from the sectors $b_1$, $b_2$ and $b_3$.
All the states at this level of the string model building are in
GUT representations. 

The second stage of the basis construction consists of adding three
additional basis vectors to the NAHE set.
These three additional basis vectors
correspond to ``Wilson lines'' in the orbifold formulation.
Three additional vectors are needed to reduce the number of generations
to three, one from each sector $b_1$, $b_2$ and $b_3$.
At the same time the additional boundary condition basis vectors
break the gauge symmetries of the NAHE set.
The $SO(10)$ symmetry is broken to one of its
subgroups $SU(5)\times U(1)$, $SO(6)\times SO(4)$ or
$SU(3)\times SU(2)\times U(1)_{B-L}\times U(1)_{T_{3_R}}$.
To break the $SO(10)$ symmetry to $SU(3)\times SU(2)\times
U(1)_C\times U(1)_L$\footnote{$U(1)_C={3\over2}U(1)_{B-L};
U(1)_L=2U(1)_{T_{3_R}}.$}
we first use a boundary condition basis vector which breaks $SO(10)$
to $SU(5)\times U(1)$ or $SO(6)\times SO(4)$. We then break the
gauge group to $SU(3)\times SU(2)\times U(1)_C\times U(1)_L$.
Since the superstring derived standard--like models contain
the $SO(6)\times SO(4)$, as well as the $SU(5)\times U(1)$,
breaking sectors, their massless spectra admits also the exotic
representations that can appear in these models.

In the superstring standard--like models,
the observable gauge group after application
of the generalized GSO projections is
$SU(3)_C\times U(1)_C\times SU(2)_L\times U(1)_L
\times U(1)^3\times U(1)^n$.
The electromagnetic charge is given by
\begin{equation}
U(1)_{\rm e.m.}=T_{3_L}+U(1)_Y,
\label{quem}
\end{equation}
where $T_{3_L}$ is the diagonal
generator of $SU(2)_L$, and $U(1)_Y$ is the weak hypercharge.
The weak hypercharge is given by{\footnote{
Note that we could have instead defined the weak hypercharge to be
$U(1)_Y={1\over 3} U(1)_C - {1\over 2} U(1)_L$. This amounts to the same
redefinition of fields between the straight and flipped $SU(5)$. In this
paper we will use the definition in Eq. \ref{qu1y}.}}
\begin{equation}
U(1)_Y={1\over 3} U(1)_C + {1\over 2} U(1)_L
\label{qu1y}
\end{equation}
and the orthogonal
combination is given by
\begin{equation}
U(1)_{Z^\prime}= U(1)_C - U(1)_L.
\label{quzp}
\end{equation}
The hidden $E_8$ gauge group is typically broken to 
$SU(5)_H\times SU(3)_H\times U(1)^2$, and the flavor
$SO(6)^3$ symmetries are broken to factors of $U(1)$s. 

The massless spectrum of the standard--like models contains
three chiral generations from the sectors which are charged under the
horizontal symmetries. Each of these consists of a 16 of
$SO(10)$, decomposed under the final $SO(10)$ subgroup as
\beqn
{e_L^c}&\equiv& ~[(1,~~{3\over2});(1,~1)]_{(~1~,~1/2~,~1)}~~;~~
{u_L^c}~\equiv ~[({\bar 3},-{1\over2});(1,-1)]_{(-2/3,1/2,-2/3)};~~~~
                                                        \label{ulc}\\
{d_L^c}&\equiv& ~[({\bar 3},-{1\over2});(1,~1)]_{(1/3,-3/2,1/3)}~~;~~
Q~\equiv ~[(3, {1\over2});(2,~0)]_{(1/6,1/2,(2/3,-1/3))};~~~~\label{q}\\
{N_L^c}&\equiv& ~[(1,~~{3\over2});(1,-1)]_{(~0~,~5/2~,~0)}~~;~~
L~\equiv ~[(1,-{3\over2});(2,~0)]_{(-1/2,-3/2,(0,1))},~~~~\label{l}
\eeqn
where we have used the notation
\begin{equation}
[(SU(3)_C\times U(1)_C);
     (SU(2)_L\times U(1)_L)]_{(Q_Y,Q_{Z^\prime},Q_{\rm e.m.})},
\label{notation}
\end{equation}
and have written the electric charge of the two
components for the doublets.

The matter states from the NS sector and the sectors $b_1$, $b_2$
and $b_3$ transform only under the observable gauge group.
In the realistic free fermionic models, there is typically one
additional sector that produces matter states transforming
only under the observable gauge group \cite{SLM}. These
states complete the representations that we identify with
possible representations of the Standard Model. In addition
to the Standard Model states, semi--realistic superstring
models may contain additional multiplets, in the $16$ and $\overline{16}$
representation of $SO(10)$, in the vectorial $10$ representation of $SO(10)$,
or the $27$ and $\overline{27}$ of $E_6$. Such states can pair up
to form super-massive states. They can mix with, and decay into,
the Standard Model representation unless some additional symmetry,
which forbids their decay, is imposed. For example, in the
flipped $SU(5)$ superstring models \cite{revamp}, two of the
additional vectors which extend the NAHE set produce an additional
$16$ and $\overline{16}$ representation of $SO(10)$. These states
are used in the flipped $SU(5)$ model to break the $SU(5)\times
U(1)$ symmetry to $SU(3)\times SU(2)\times U(1)$.

In addition to the states mentioned above transforming solely
under the observable gauge group, the sectors $b_j+2\gamma$
produce matter states that fall into the 16 representation of the hidden
$SO(16)$ gauge group decomposed under the final hidden gauge group.
The states from the sectors $b_j+2\gamma$ are $SO(10)$ singlets, but
are charged under the flavor $U(1)$ symmetries. All the states above
fit into standard representation of the grand unified group which may be,
for example, $SO(10)$ or $E_6$, or are singlets of these groups.
They carry the standard charges under the Standard Model gauge
group or of its GUT extensions. The superstring models, however,
contain additional states that cannot fit into multiplets of the
original unifying gauge group. Such states are exotic stringy states and cannot
fit into representations of the underlying $SO(10)$ symmetry group of
the NAHE set. They result from the breaking of the $SO(10)$ gauge
group at the string level via the boundary condition assignment. 
In the next section
we enumerate the states that appear in free fermionic models.

\section{Exotic matter}

The boundary condition basis vectors, beyond the NAHE set,
are used to break the unifying $SO(10)$ gauge group.
Since these sectors correspond to ``Wilson lines'' they give rise
to massless states that do not fit into representations of the
original $SO(10)$ symmetry. As a result the massless spectrum contains
states with fractional charges under the unbroken $U(1)$ generators
of the original $SO(10)$ gauge group. This is a new and common feature
of superstring models. In many examples the exotic states appear in
vector--like representations and can acquire a heavy mass.
The ``Wilsonian'' matter phenomenon is an important feature
as it may result in discrete symmetries that may prevent the
decay of the exotic massive states into the Standard Model states,
and therefore give rise to dark matter candidates.
The following exotic matter representations can appear in free
fermionic level one models. The states are classified according
to the unbroken $SO(10)$ subgroup in each sector \cite{ccf,SLM}.

{}From the $\underline{SO(6)\times SO(4)}$ type sectors we obtain the
following exotic states.

\bigskip
$\bullet~ {\underline {\rm Color ~triplets:}}~~~~
  [(    3, {1\over2});(1,0)]_{( 1/6, 1/2, 1/6)}~~~~;~~~~
  [({\bar3},-{1\over2});(1,0)]_{(-1/6,-1/2,-1/6)}$
\parindent=15pt
\bigskip

$\bullet~{\underline{\rm Electroweak ~doublets:}}~~~~
[(1,0);(2,0)]_{(0,0,\pm1/2)}$

\bigskip

$\bullet$ {\underline{Fractionally charged
                $SU(3)_C\times SU(2)_L$ singlets :}}
\beq
[(1,0);(1,\pm{1})]_{(\pm1/2,\mp1/2,\pm1/2)}~~~~;~~~~
[(1,\pm3/2);(1,0)]_{(\pm1/2,\pm1/2,\pm1/2)}
\label{fc64singlet}
\eeq
\parindent=15pt
The color triplets bind with light quarks to form mesons and baryons
with  fractional electric charges $\pm1/2$ and $\pm3/2$.
The $\underline{SO(6)\times SO(4)}$ type states can appear 
in the Pati--Salam type models \cite{ALR} or in Standard--like models.

{}From sectors which break the $SO(10)$ symmetry into 
$\underline{SU(5)\times U(1)}$ we obtain exotic states
with fractional electric charge $\pm1/2$

\bigskip

$\bullet$ {\underline{Fractionally charged $SU(3)_C\times SU(2)_L$ singlets :}}
\beq
[(1,\pm3/4);(1,\pm{1/2})]_{(\pm1/2,\pm1/4,\pm1/2)}
\label{fc51singlet}
\eeq
\parindent=15pt

In general the fractionally charged states may transform
under a non--Abelian hidden gauge group in which case the fractionally
charged states may be confined.
For example, in the ``revamped'' flipped $SU(5)$ model \cite{revamp}
the states with fractional charge $\pm1/2$ transform as $4$ and $\bar4$
of the hidden $SU(4)$ gauge group. The states with the charges in
eq. (\ref{fc51singlet}) are called the ``cryptons'' and
may form good dark matter candidates \cite{eln}
if the lightest confined state is electrically neutral.
In the ``revamped'' flipped $SU(5)$ model it has been
argued that the lightest state is the ``tetron'',
which contains four fundamental constituents. 
In other models, states with the charges of eq. (\ref{fc51singlet})
may be singlets of all the non--Abelian group factors.

Finally in the superstring derived standard--like models
we may obtain exotic states from sectors which are combinations
of the $\underline{SO(6)\times SO(4)}$ breaking vectors and 
$\underline{SU(5)\times U(1)}$
breaking vectors. These states therefore arise only in
the $\underline{SU(3)\times SU(2)\times U(1)^2}$ type models. 
These states then carry the standard charges
under the Standard Model gauge group but carry fractional charges
under the $U(1)_{Z^\prime}$ gauge group.
The following exotic states are obtained:

\bigskip

$\bullet~ {\underline{\rm color ~triplets :}}$
\beq
[(3,{1\over4});(1,{1\over2})]_{(-1/3,-1/4,-1/3)}~~~~;~~~~
[(\bar3,-{1\over4});(1,{1\over2})]_{(1/3,1/4,1/3)}
\label{unit}
\eeq

\bigskip
\parindent=15pt

Due to its potential role in string gauge coupling unification
\cite{gcu}, this state is referred to as ``the uniton'' \cite{ccf}.
The uniton forms bound states with the lightest up and down
quarks and gives rise to ultra--heavy mesons. In ref. \cite{ccf}
it has been shown that the lightest meson can be the electrically
neutral state. 

\bigskip

$\bullet~ {\underline{\rm electroweak ~doublets : }}~~~~
[(1,\pm{3\over4});(2,\pm{1\over2})]_{(\pm1/2,\pm1/4,(1,0);(0,-1))}$

\bigskip
\parindent=15pt

Unlike the previous electroweak doublets, these electroweak doublets
carry the regular charges under the standard model gauge group but carry
``fractional'' charge under the $U(1)_{Z^\prime}$ symmetry. 
Finally, in the superstring derived standard--like models we also obtain
states which are Standard Model singlets but carry ``fractional''
charges under the $U(1)_{Z^\prime}$ symmetry.

\bigskip

$\bullet$ {\underline{Standard model singlets with ``fractional''
                        $U(1)_{Z^\prime}$ charge :}}

\beq
[(1,\pm{3\over4});(1,\mp{1\over2})]_{(0,\pm5/4,0)}
\label{fc321singlet}
\ee
\parindent=15pt
These states may transform under a non--Abelian hidden gauge group
or may be singlets of all the non--Abelian group factors.
This type of Standard Model singlet appears in all the
known free fermionic standard--like models.

There are several important issues to examine
with regard to the exotic states. Since some of these states carry
fractional charges, it is desirable to make them sufficiently heavy
or sufficiently rare. A priori, in a generic string model, it is not
at all guaranteed that the states with fractional electric charge can
be decoupled or confined \cite{lykken}. 
Therefore, their presence imposes an highly
non--trivial constraint on potentially viable string vacua. 
In the NAHE--based free fermionic models, all the exotic
matter states appear in vector--like representations. They can therefore
obtain mass terms from renormalizable or higher order terms in the
superpotential. We must then study the renormalizable and
nonrenormalizable superpotential in the specific models.
The cubic level and higher order terms in the superpotential are
extracted by applying the rules of ref. \cite{KLN}.
The problem of fractionally charged states
was investigated in ref. \cite{fcp,cfn} for the model of ref.
\cite{fny}. 
By examining the fractionally charged states and the trilinear
superpotential, it is observed that all the fractionally charged
states receive a Planck scale mass by giving a VEV to
the neutral singlets
${\bar\phi}_4,{\bar\phi}_4',{\phi}_4,\phi_4'$
which imposes  the additional F flatness constraint
$(\phi_4{\bar\phi}_4'+{\bar\phi}_4\phi_4')=0$. The other exotic states
which are Standard Model singlets do not receive mass by this choice
of flat direction. Therefore, at this level of the superpotential, the
fractionally charged states can decouple from the remaining light
spectrum. Similarly, the issue of fractionally charged states in the
model of ref. \cite{eu} was studied in ref. \cite{nrt} where it was
found that all the fractionally charged states receive large mass from
renormalizable or nonrenormalizable terms. Similar results
were also found in the case of the Gepner models \cite{huet}.
The second issue that must be examined with regard to the
exotic ``Wilsonian'' matter is the interactions with the Standard Model
states. The fractional charges of the exotic states
under the unbroken $U(1)$ generators of the $SO(10)$ gauge group,
may result in conserved discrete symmetry which forbid their
decay to the lighter Standard Model states \cite{ccf}.

\section{Ultra energetic neutrinos from the sun and the earth}

We next examine the predicted flux of energetic 
neutrinos from the sun and the earth for several exotic stringy states 
as well as other meta--stable states, which have been proposed in
the literature. The exotic states which we study are : 
1) The ``uniton'' from eq. (\ref{unit}); 2) the Gluino--LSP of ref.
\cite{raby}; 3) the ``crypton'' from eq. (\ref{fc51singlet});  4) the
exotic Standard Model singlet from eq. (\ref{fc321singlet}).

Before computing the detailed neutrino fluxes, it will be useful to discuss
the generic features of the indirect dark matter detection with superheavy
particles of mass
$M_X$.  General scaling properties can be established and will
depend  mainly on the type of forces inducing the interaction of 
dark matter particles $X$ with matter. Ultimately, for every type of force, 
there is a maximal mass $M_X$ which can be probed by annihilation in the
sun or the earth.

We concentrate our analysis on the trapping and annihilation of 
heavy particles inside the cores of the sun and the earth \cite{earth},
and the detection of the energetic neutrinos by underground detectors. 
 Another, very important issue is the relation between 
direct and indirect searches for ultra heavy particles. 
If the density of massive particles in the galactic halo is fixed 
by the dark matter density ($\simeq 0.3$ GeV cm$^{-3}$), then the scaling
of the event rate for  the direct search of a dark matter candidate is
proportional to $\sigma/M_X$ which appears to be more  
advantageous than the $1/M_X^2$ fall off of the trapping rate. For large
$M_X$, direct detection is based on registering the recoil energy of the
nuclei during the  elastic collision with the dark matter particle. Since
the recoil energy is not sensitive to the scale $M_X$, when $X$
is much heavier then the nucleus, direct detection
does not allow one to gain sensitivity due to the increasing 
mass of the particle. On the
other hand, the probability, $P$, for the conversion of neutrinos to
muons does increase rapidly with mass (for masses $\ga 100$ GeV, the
increase is stronger than linear), and indirect detection is more
sensitive to  the presence of dark matter.  At very high masses, the
conversion probability flattens out and eventually direct detection
wins. We will discuss the probability function
$P(M_X)$ in more detail below. Here, we will restrict our attention to
indirect detection by annihilation in the cores of the sun and the earth.

Energetic 
neutrinos from the sun and the earth  have already been discussed
as a possible signal for neutralino dark matter \cite{sos,kamion,berg}.
Therefore, from the experimental perspective
there are two relevant issues. 
The first is the expected flux of energetic
neutrinos for each dark matter candidate and
the feasibility of observing it in contemporary 
detectors. The second issue is the experimental
prospect of distinguishing among the various
dark matter candidates. The energetic neutrinos
are detected by observing the muons that they produce 
upon interacting with matter in the vicinity
of the detector. Currently the experiments
can only determine the flux of muons above a certain
cutoff. Therefore, the prospect for experimentally
distinguishing among the candidates depends on
the ability of experiments to record the actual
energy of the muon or to apply a variable energy threshold. 

The mechanisms responsible for the production of very energetic neutrinos 
inside the earth and the sun are very similar. In what follows 
we concentrate in details on the trapping and annihilation of
heavy particles 
in the core of the sun and simply present similar estimates for the earth. 
Ultra energetic neutrinos from
the sun arise if the ultra heavy particles from the 
galactic halo are trapped in the sun and subsequently
annihilate into neutrinos. The expected flux
depends on the competition between the trapping and annihilation rates. 
Following ref. \cite{sos}, we express the initial 
density, $n_X$, of ultra heavy particles of mass, $M_X$, 
in the halo as fraction of the required dark matter density
\be
M_Xn_X= \delta_X\bar v^2/6\pi G a^2,
\ee
where $\bar v\simeq 300$ km/s is the r.m.s. velocity of a halo $X$ in 
the solar neighborhood and $a = 10$ kpc. The coefficient $\delta_X\leq 1$
is kept as a free parameter, reflecting the partial contribution of 
particles $X$ into the total dark matter density. 
 
The elastic scattering on nuclei, $X+N~\rightarrow~ X+N$, 
leads to the gravitational trapping of $X$ particles inside the sun. 
The trapping rate is given by
\beq
\Gamma_T =(7.3 \times 10^{28}{\rm sec}^{-1})
\delta_X{1 {\rm GeV}\over{M_X}}\sum_N
\sigma_{N,36}Y_Nf_{E,N},
\label{traprate}
\eeq
where
$\sigma_{N,36}$ is the elastic cross section in units of $10^{-36}$cm$^2$ 
and $Y_N$ is fraction of nuclei $N$ by number in the sun. 
The most important factor here is $f_{E,N}$, 
the kinematic parameter, accounting for the fraction of the halo 
particles which loose enough energy to get trapped.  
In the limit of very heavy $M_X$, $f$ reduces to \cite{ps}
\beq
f_{E,N}\simeq\fr{43 m_N}{M_X}
\label{fe}
\eeq 
Therefore, the scaling of the trapping rate with $M_X$ (for very large
$M_X$) goes as 
\beq
\Gamma_T\sim {\sigma_{\rm elastic}\over{M_X^2}}
\eeq
If the cross section is constant the trapping rate exhibits a quadratic
fall off behavior. 

Another important quantity, determining the flux of energetic neutrinos
is the annihilation rate given by \cite{sos}
\beq
\Gamma_A=(5 \times
10^{54}s^{-1})\left({{n_{X_{\odot}}m_P}\over\rho_\odot}\right)^2
\langle\sigma v\rangle_{A,26}W.
\eeq
Here $\langle\sigma v\rangle_{A,26}=
\langle\sigma v\rangle_{A}/10^{-26}{\rm cm}^3{\rm s^{-1}}$
is the low--energy annihilation cross section
times relative velocity, $n_{X_{\odot}}$ is the mean
$X$ particle density inside the sun, $\rho_\odot$
is the mean mass density of the sun. $W$ is the 
result of the non-uniform distribution of $X$ particles inside the sun,
\be
W=\fr{\int n_{X_{\odot}}^2(r)dV}{n_{X_{\odot}}^2V_\odot}=
V_\odot\fr{\int n_{X_{\odot}}^2(r)dV}{(\int n_{X_{\odot}}(r)dV)^2}
\label{w}
\ee
{}From ref. \cite{grs}, we take
\be
W\simeq 250\left(\fr{M_X}{1{\rm GeV}}\right)^{3/2}
\ee
The annihilation rate therefore scales as
$(n_{X_{\odot}})^2 /(M_X)^{1/2}$ assuming a natural scaling 
$\langle\sigma v\rangle\sim 1/M_X^2$.

Usually, the annihilation rate is a much slower process if the density of 
$X$ particles inside the sun $n_{X_{\odot}}$ is of the order of their density 
$n_X$ in the halo. This situation is unstable, leading to the accumulation 
of dark matter candidates in the sun 
up to the level when the annihilation and trapping rates become equal
\beq
\Gamma_A=\Gamma_T
\label{equil}
\eeq
However, for large masses, the time needed to reach the equilibrium
condition,
\beq
\tau_{eq}= (n_{X\odot}M_\odot/\rho_\odot)/\Gamma_T\simeq 2\times
10^{13}{\rm sec}~
\delta_X^{-1/2}\left(\fr{M_X}{1{\rm GeV}}\right)^{5/4}\fr{1{\rm
GeV}}{\sqrt {M_X^2\langle\sigma v\rangle_{A,26}\sigma_{p,36}}}
\ee
may turn out to be larger than the age of the sun $\tau_\odot$. 
For a constant elastic cross section, and an annihilation cross
section which drops off as $M_X^{-2}$, this quantity scales as
$\tau_{eq}\sim  M_X^{5/4}$.
The condition $\tau_{eq}<\tau_{\odot}$ must be checked for every 
particular dark matter candidate. 

If equilibrium is reached, the flux of neutrinos from the 
annihilation inside the core of the sun is determined 
by the trapping rate and by
an average number of neutrinos $N_{eff}$,
\begin{eqnarray}
\phi_{\nu\odot}=\fr{1}{2}\Gamma_TN_{eff}/4\pi(1\mbox{A.U.})^2
\left\{
\begin{array}{c}
1~~~~~~ {\rm for}~~\tau_\odot\gg\tau_{eq}\\
 (\tau_\odot/\tau_{eq})^2 ~~~~~~ {\rm for}~~\tau_\odot\ll\tau_{eq}
\end{array}\right.\simeq\nonumber\\ (560 cm^{-2}s^{-1})
N_{eff}\delta_X\sigma_{p,36}\fr{\mbox{GeV}^2}{M_X^2}
\left\{
\begin{array}{c}
1~~~~~~ {\rm for}~~\tau_\odot\gg\tau_{eq}\\
 (\tau_\odot/\tau_{eq})^2 ~~~~~~ {\rm for}~~\tau_\odot\ll\tau_{eq}
\end{array}\right.,
\label{flux}
\end{eqnarray}
where we have kept only the contribution of the proton in the elastic
cross section. $N_{eff}$ should be interpreted as an everage number of 
neutrinos {\em escaping} from the surface of the sun and produced 
as a consequence of a single annihilation event inside the core.

Using very similar assumptions our estimate for the flux of the neutrinos 
generated in the center of the earth is 
\begin{eqnarray}
\phi_{\nu\oplus}=\fr{1}{2}\Gamma_TN_{eff}/4\pi R_\oplus^2
\left\{
\begin{array}{c}
1~~~~~~ {\rm for}~~\tau_\oplus\gg\tau_{eq}\\
 (\tau_\oplus/\tau_{eq})^2 ~~~~~~ {\rm for}~~\tau_\oplus\ll\tau_{eq}
\end{array}\right.\simeq\nonumber\\ (0.05 cm^{-2}s^{-1})
N_{eff}\delta_X\sum_NY_N\sigma_{N,36}\fr{\mbox{GeV}^2}{M_X^2}
\left\{
\begin{array}{c}
1~~~~~~ {\rm for}~~\tau_\oplus\gg\tau_{eq}\\
 (\tau_\oplus/\tau_{eq})^2 ~~~~~~ {\rm for}~~\tau_\oplus\ll\tau_{eq}
\end{array}\right.
\label{flux1}
\end{eqnarray}
where, of course, $\tau_{eq}$ and $N_{eff}$ must be adjusted from those in
eq. (\ref{flux}).

A more precise treatment cannot be achieved in a model 
independent way and therefore, we next turn to specific candidates.

\section{High energy neutrinos produced by superstring relics }

The initial flux of neutrinos produced close to the center of the sun 
or the earth can be estimated relatively easily. 
This flux
must contain a significant portion of neutrinos with energies $\sim M_X$. 
The subsequent fate of highly energetic neutrinos is very different and 
very model and energy dependent. In the formulae for neutrino fluxes
(\ref{flux}--\ref{flux1}), 
$N_{eff}$ is the number of neutrinos produced in a
single annihilation event times the probability for high energy neutrinos 
to reach the vicinity of a detector. 

In general, an $X\bar X$--pair can annihilate into various SM particles
including neutrinos. If neutrinos are the direct product of the decay, 
we should expect to have 
significant absorption of the neutrinos with energies larger than few
hundred GeV \cite{Ni} in the core of the
sun. This occurs due to the conversion of neutrinos
into muons as the result of charged current 
interactions and leads to the exponential loss of the 
signal as muons are efficiently 
stopped by the solar media and most of them decay at rest. 
It does not mean, however,
that the effective number of the neutrinos escaping from the surface of 
the sun is negligibly small even though the absorption length is smaller 
than the radius of the sun. Other annihilation products of an $X\bar
X$--pair will contribute to $N_{eff}$. In fact, the hadronic products of 
$X\bar X$-annihilation can be an important source of neutrinos. 
The decay of hadrons containing different
flavors lead to a broad spectrum of neutrinos as hadronic lifetimes are 
significantly different. The average energy
of the neutrinos produced in the decay of heavy quarks except for
the $t$--quark is much smaller than $M_X$ and determined to be of order a
few hundred GeV 
\cite{Ni}. For these neutrinos, absorption inside the sun does not 
lead to a drammatic attenuation of the signal. This means that in the limit 
$M_X \gg 1$ TeV, the hadronic decay channels are mostly responsible for
the  production of neutrinos which can reach the surface of the sun
without being  converted into muons. Inside the Earth the absorption
of the neutrinos is not the problem as the attenuation of the signal is
very small up to very  high energies, $M_X\sim 10^9$ GeV \cite{PTP}.

For our numerical estimates we will assume the following 
crude pattern for the spectrum. The flux of the neutrinos 
coming from the sun originating
from the decay of $X$-particles with $M_X\gg 1$ TeV is downgraded in the solar
media by approximately two orders of magnitude and the maximum of 
neutrino energy distribution is of the order of few hundred GeV. If the 
energy of the annhilating particles is less than 1 TeV, $N_{eff}$ can be larger, 
$N_{eff}\sim O(1)$, especially if the direct channels of 
decay into neutrinos are open. For the earth we assume that the spectrum of
neutrinos reaching the vicinity of a detector is basically the same as 
it is produced near the center and the fraction of
very energetic neutrinos is not suppressed. 
 
The most important charactersitic, determining the flux of neutrinos 
is the elastic cross section, as it can vary by orders of magnitude
and even scale differently with $M_X$ for different candidate choices.
The behavior of the elastic cross section on ordinary matter for the 
different types of dark matter candidates we consider (stable
string relics, fractionally charged states etc.)  
depends on several generic features of these particles. Let us assume
that heavy particles $X$ does  not carry an open charge with respect
to SM gauge group but is composed of constituents which are
charged with respect to standard
$SU(3)\times SU(2)\times U(1)$. The confining force may be the usual 
strong/electromagnetic interaction or a new gauge interaction 
as is the case for ``cryptons''. 
The size of the wave function for such a system will be
determined by the {\em lightest} constituent and the nature of the
confining force. If all constituents are heavy, of the order $M_X$, 
then the interaction between such a particle and normal matter will 
at least scale as $1/M_X$ and the cross section will fall off as
$1/M_X^2$. The presence of a ``light component'' inside this particle 
(light quark, electrons etc.) will lead to a constant cross section,
which does not vanish in the limit, $M_X\rightarrow \infty$. 

The first case we consider here is the uniton.
A large neutrino flux from the sun can arise in the case of 
uniton, as its interaction with ordinary matter is mediated by 
strong interactions.
Depending  on the type of uniton 
(its total spin, internal quantum numbers, etc.) the
interaction with ordinary matter will be mediated by 
$\pi$, $\sigma$, $\rho$ mesons. The natural estimate for the cross
section is
\be
\sigma_{(uniton, N)}\sim (200\mbox{MeV})^{-2}\sim 10^{-26}{\rm cm}^2.
\label{unitonel}
\ee
The independence of the elastic cross section with respect to $M_X$ in the
case of  uniton is obvious. The heavy component of the exotic hadron ($X$
particle) participates in the scattering only as a spectator. 
The interaction is determined by the light component(s) of the exotic hadron 
wave function and its strength is set by the its size and by the energies 
of the excited states, that is by the quantities related to $\Lambda_{QCD}$.
The independence of the elastic cross section from $M_X$ can be understood
from the simple analogy with cross section of scattering of heavy atoms at
low energies. The latter is determined by the size of the atoms, the wave
functions of outer electrons, atomic polarizabilities, etc. and does not 
depend on the masses of the heavy nuclei.  

For the annihilation cross section we take $\langle\sigma
v\rangle_{A}\sim 
\alpha_s^2/M_X^2$. The time needed to reach the equilibrium
density of unitons inside the sun is
\be
\tau_{eq,uniton}\sim 6\times 10^4{\rm sec}~ \delta_X^{-1/2}
\left(\fr{M_X}{1\mbox{GeV}}\right)^{5/4},
\ee
If $\delta_X\sim 1$, equilibrium is reached so long as $M_X \la 10^{10}
GeV$. For $X$-particles trapped inside the earth the time 
to reach equilibrium is $\sim 5\cdot 10^4$ times larger 
which can be realized for $M_X \la 10^{6} GeV$.

Assuming equilibrium, we can easily obtain the neutrino flux by
plugging the estimate (\ref{unitonel}) into the expression for the
neutrino flux (\ref{flux})
\begin{eqnarray}
\phi_{\nu\odot} \sim 5.6\times 10^{12} 
{\rm cm}^{-2}{\rm s}^{-1} N_{eff}\delta_X 
\left(\fr{1\mbox{GeV}}{M_X}\right)^2\nonumber\\
\phi_{\nu\oplus} \sim 5\times 10^{8} 
{\rm cm}^{-2}{\rm s}^{-1} N_{eff}\delta_X 
\left(\fr{1\mbox{GeV}}{M_X}\right)^2,
\end{eqnarray}
If $N_{eff}\sim O(10^{-2})$ and $\delta_X\sim O(1)$ these fluxes are 
enormous, which results in 
strong constraints on the mass and/or density of unitons. To compare
it with existing experimental data we must include 
the effectiveness of the conversion 
of highly energetic muon neutrinos into muons. The flux of the 
muons can be calculated \cite{GS,PTP} by including the probability $P$
for a neutrino to convert into a muon and the muon reaching the
detector.  This probability rises as $E^2$ for energies
below $\sim$ 1 TeV.   At 1 TeV, 
$P(E=1\mbox{TeV})\sim10^{-6}$ and flattens at higher energies 
\cite{GS,PTP}. The actual value of $P$ at high energies
differs in these two works but for our purposes it is reasonable to 
take $P(E>10^7 {\rm GeV})\sim 10^{-3}$, which is a conservative 
estimate since the probability is certainly higher at higher energies.

Comparing the resulting flux of the energetic muons with the existing bounds
from the KAMIOKANDE, MACRO and BAKSAN collaborations, $\phi_\mu \la
1.4\times  10^{-14}$ cm$^{-2}$s$^{-1}$ \cite{Kam,Mac,Bak}, and assuming a
maximal abundance of unitons and, we deduce the bound on the
uniton mass $M_X$:
\beq
M_X > 10^{9}-10^{10} \mbox{GeV}
\label{constr}
\ee
This bound is obtained from the sun generated flux, assuming 
$N_{eff}\sim 0.01$ and $P\sim 10^{-6}$
and it is roughly consistent with the equilibrium conditions.
The earth generated signal is 
less restrictive largely because the equilibrium condition is violated for
$M_X> 10^6$ GeV.

In contrast to eq. (\ref{constr}), bounds on the uniton 
mass from its relic density due to
annihilation in the early Universe require $M_X < 10^7 -10^8$ GeV
\cite{ccf}.  Our constraint
(\ref{constr}) therefore, rules out the uniton as a dark matter
candidate.  It is interesting to note that the bound (\ref{constr})
is similar to the mass range required for a superheavy relic decay to
produce the ultra-high energy cosmic-rays.  Thus, if a strongly
interacting, very massive, dark matter particle is the source of the
ultra-high energy cosmic-rays, then a signal in the form of neutrinos
from the sun and the earth should arise as the indirect detection experiments
improve their sensitivity.

If we relax the constraint on the mass by making the cosmological 
abundance, $\delta_X$, small, we can use estimates similar to the ones
made above for the gluino-LSP scenario which has been
advocated recently in \cite{raby}.
In this case, the cosmological abundance of gluino--containing hadrons
was estimated to be at the level of $\delta_X\sim 10^{-9}(M_X/1 {\rm
GeV})$.  The cross-section of the gluino-hadron with ordinary matter
is quite large and can be estimated to be
$10^{-26}$ cm$^2$ as in the case with uniton. For a gluino mass of
order $100$ GeV, this abundance  would lead to a muon flux from the sun
five orders of magnitude above current  experimental limit, even if we
assume a low value for $N_{eff}\sim 0.01$. Thus, here too, the indirect
detection limits provide a strong constraint and are complementary to
the astophysical constraints discussed in \cite{mt}.  The gluino-LSP bound
state is forbidden to roam freely in the galactic halo and as such can not
provide a source for the ultra-high energy cosmic-rays unless it is
infact unstable and therefore not the LSP (or R-parity is broken). As we
noted earlier, our constraints do not apply to the light gluino scenario
proposed in
\cite{cfk}, since the neutrino energies will be below threshold for the
existing detectors. 

Let us consider now the case of ``cryptons'', neutral objects 
composed from the constituents which bear $U(1)_{em}$ charge (``cryptons'')
held together by a confining force of non--SM origin.
It is clear that their interaction with ordinary matter is suppressed
because of the absence of light constituents  in their structure.
The absence of 
an overall charge with 
respect to the electromagnetic gauge group does not preclude these particle 
from possessing an anomalous magnetic moment provided that they have 
spin and $X$ and $\bar X$ are different.
The size of the anomalous magnetic moment depends on the details
of the force confining the charged constituents together. It is natural
to assume, however, that the size of the anomalous magnetic moment is
similar to an ``exotic Bohr magneton'', $e/(2M_X)$. 
In this case the fall off of the cross section with $M_X$ is just quadratic,
\beq
\sigma_p\sim \fr{\alpha^2}{M_X^2}
\label{crysect}
\ee
Still, this decrease is very sharp (especially if we are interested in
very massive states) and the corresponding neutrino flux is low.
Inserting (\ref{crysect}) into the neutrino flux from the sun (\ref{flux}),
we find
\be
\phi_{\nu\odot} \sim 2.2\times 10^{11} {\rm cm}^{-2}{\rm s}^{-1} \alpha^2
N_{eff}\delta_X 
\left(\fr{1\mbox{GeV}}{M_X}\right)^4.
\ee
The earth signal is additionally suppressed as most of the nuclei inside 
the earth are spinless.
Normalizing the probability of conversion to muons at the neutrino energy 
scale of 1 TeV, 
$P=10^{-6}P_6$, we find the following constraint 
\be
N_{eff}\delta_XP_6\left(\fr{1 {\rm TeV}}{M_X}\right)^4 
< 1.2 \times 10^{-3}
\label{cryp}
\ee
Equilibrium is reached for $M_X$ lighter than a few TeV, if we take 
annihilation and elastic scattering cross sections to be of the same order,
$\langle \sigma_Av\rangle\sim \sigma_p$. Though (\ref{cryp}) is
significantly  weaker than the bound on the uniton, it is not altogether
trivial.

The last case we consider is that of the pure SM singlets, charged with
respect to an extra
$U(1)$ group. The interaction of these particles with ordinary matter
will depend on the mass of the gauge boson associated with this
additional group.  The size of the elastic cross section mediated by
exotic $Z'$ can be estimated as
\beq
\sigma_p\sim \fr{g^{'2}_X}{4\pi}\fr {g^{'2}_p}{4\pi}\fr{ m_p^2}{M_{Z'}^4}.
\ee
Here $g'_X$ and $g'_p$ denote total charges of exotic and ordinary 
matter with respect to this additional $U(1)$ group. 
If $Z'$ is very heavy, of the order $M_X$, then this cross section is very 
small. $M_{Z'}$ can be kept as a free parameter, not necessarily related to
$M_X$. However, direct searches for $Z'$'s place a lower limit on its mass
at the level  of 500 $GeV$, which is still too heavy to produce a large
elastic cross section. Assuming $M_{Z'}<M_X$, we obtain the following
constraint on  on the mass of $Z'$ and the combination of charges,
\be
N_{eff}\delta_XP_6\fr{g^{'2}_Xg^{'2}_p}{e^4}
\fr{1 {\rm TeV}^6}{M_{Z'}^4M_X^2} < 1.2 \times 10^{-3}
\ee

Numerically, the limits derived for the two cases, ``cryptons'' and SM
singlets with
$Z'$--charge, are similar to those placed on a conventional
neutralino  dark matter scenario, suggesting that the current 
sensitivity to the masses of these
particles cannot be much better than few $TeV$. 

\section{Discussions and conclusions}

Perhaps more exciting than setting a limit on exotic states is the
possibility of detecting a high energy neutrino signal from the sun
or the earth.
The detection of an ultra high energy neutrino from the earth or the sun would in
fact imply a lower limit to the fundamental cut--off scale. If the origin of
the neutrino can be correlated with the position of the sun or with the 
direction to the center of the earth, one would
be forced to conclude that such a neutrino could only be produced via
the annihilation (or decay) of a massive particle with $m \sim
E_\nu$. In models with large extra dimensions, the fundamental cut--off
scale is necessarily below the 4-dimensional Planck scale as
given by
$$ L^n M^{2+n} = M_P^2$$
for the compactification of a $4+n$ dimensional theory where $M$
is the fundamental cut--off scale and $L$ is the characteristic size of
the extra dimensions.  It is clear that in such a model, the
maximum particle mass is also
$M$. For example, in the $4+n$ dimensional theory, the maximum
mass of a scalar particle is $m
\sim M$ (above $M$, the full quantum theory would be required).
The scalar mass term $m^2 {\hat \phi}^2$ where ${\hat \phi}$ is
the $4 + n$-dimensional scalar, becomes $m^2 \phi^2$, where
$\phi = L^n {\hat \phi}$ after the scalar kinetic term is
renormalized. Thus the limit $m \le M$ is maintained in four
dimensions.  The limit also applies to the KK excitations of a
lighter field. In 4-dimensions, the KK masses of the
excitations scale as $m_{KK} \sim n/L$.  However the number of
such states is limited by the maximum momentum in the extra
dimension which is $M$.  Therefore, the highest mass state we
can consider is $N/L \sim M$.  The same argument holds for the
maximum mass of a fermion as well.  Therefore, any detection of
a high energy neutrino for which we can be sure was produced in
the decay or annihilation of a  particle with an ultra heavy
mass $M_X$, sets a firm lower bound on the fundamental scale,
$M \ge M_X$.

If we are interested in setting a strong constraint on the fundamental
cut--off scale (that is one which is significantly stronger than the current
experimental limit of $M \ga 1$ TeV), a detector with sufficiently
high angular resolution is needed. The large water detectors such as
Kamiokande have an angular resolution of order $1\deg$. In this case,
energies above 10's of GeV can not be distinguished and an interesting
constraint on the fundamental cut--off scale can not be set.  In contrast, a
detector such as Soudan II, which has an angular resolution of order
0.1\deg, will be able to correlate only very energetic ($E \gg 1$ TeV)
neutrinos with the centers of the sun or the earth.  Of course, such a constraint
requires a positive detection of ultra-high energy neutrinos from the
sun and/or the earth. 

\medskip

In summary, we have argued that the search for the highly energetic
neutrinos coming from the annihilation of very massive particles inside
the sun and the earth can be used to probe energy ranges above the
electroweak scale.

The flux of highly
energetic neutrinos is crucial to the indirect detection of dark matter. 
The flux has a universal scaling behavior, $1/M^2_X$, if the magnitude of
the elastic cross section is constant. However, the actual fall off (with
$M_X$) of the induced muon flux is much milder, since the probability of 
the conversion of neutrinos to muons rises. The elastic cross section with
ordinary matter is of course candidate specific, and its magnitude depends
on the  nature of forces mediating the interaction and on the presence 
of a ``light component'' in the wave function of the exotic state. 

In the case of the uniton and gluino-LSP, there are strong
interactions which induce  the elastic cross section with the ordinary
matter and the exotic particle has a light component $u$, $d$ or
$A_\mu^a$ in it. As a result, the  trapping rate is large leading to very
large neutrino signal unless  the the mass of $X$ particle is bigger than
$10^{9}$ GeV or their initial  density is much smaller than one needs
for dark matter energy density. In the case of the gluino-LSP candidate,
even the low  density expected for these particles does not preclude a
large detection rate of neutrinos from the sun.

We believe that the dedicated search for ultra-high energy neutrinos
coming from the sun can be very important in setting a lower bound on the 
``fundamental scale''. A signal, correlated with the positions of the sun
and/or the center of the earth,
can only be the consequence of the decay or annihilation of heavy
particles inside the sun/earth. At $M_X > 1$ TeV, neutrinos from the 
center of the earth will be more energetic and a significant fraction of
the  spectrum will have energies of order $M_X$. In contrast, the maximum
of  the neutrino spectrum coming from the sun appears to be shifted
towards  energies less than 1 TeV. 
Once the neutrino signal is identified, one can  use
the shape of the energy spectrum of the muons and try
to reconstruct the value of the mass for these particles which can be
well above the reach of accelerator physics. The existence of
heavy particles with masses
$M_X$ would imply that the fundamental (string/quantum gravity) scale is
at least as large as $M_X$. It may serve as a new way of constraining
scenarios with large extra dimensions and low (as low as 1 TeV)
fundamental Plank scale. However, this constraint does require a positive
signal rather than a flux limit.

\bigskip
\medskip
\leftline{\large\bf Acknowledgments}
\medskip

We are pleased to thank Francis Halzen, Marvin Marshak, Earl Peterson, and 
Adam Ritz for very helpful discussions. We thank  
Lars Bergstrom and Joakim Edsjo for pointing out 
the importance of the neutrino absorption inside the sun.

This work was supported in part by the Department of Energy
under Grant No.\ DE-FG-02-94-ER-40823.



\bibliographystyle{unsrt}

\end{document}